\documentclass[preprint,aps,tightenlines,showkeys,nofootinbib,superscriptaddress]{revtex4}
\usepackage{latexsym}
\usepackage[dvips]{color}
\usepackage[dvips]{graphicx}
\usepackage{bigints}
\usepackage{ulem}
\newcommand{\beq}{\begin{eqnarray}}
\newcommand{\eeq}{\end{eqnarray}}
\newcommand{\bea}{\begin{eqnarray*}}
\newcommand{\eea}{\end{eqnarray*}}
\newcommand{\eq}{eqnarray}

\newcommand{\De}{\Delta}
\newcommand{\ka}{\kappa}

\newcommand{\la}{{\lambda}}
\newcommand{\La}{{\Lambda}}
\newcommand{\m}{{\mu}}
\newcommand{\n}{{\nu}}
\newcommand{\si}{{\sigma}}
\newcommand{\Si}{{\Sigma}}

\newcommand{\pa}{{\partial}}

\newcommand{\f}{\frac}
\newcommand{\ra}{\rightarrow}

\newcommand{\Sch}{Schwarzschild }

\newcommand{\DiffF}{${\it Diff}_{\cal F}$}

\newcommand{\D}{{\rm d}}

\newcommand{\bc}[1]{\textcolor{blue}{#1}}

\begin{document}

\preprint{arXiv:2509.19744v2[hep-th]}

\title{Chronology Protection of Rotating Black Holes\\ in a Viable Lorentz-Violating Gravity}

\author{Mu-In Park \footnote{E-mail address: muinpark@gmail.com, corresponding author}}
\affiliation{ Center for Quantum Spacetime, Sogang University,
Seoul, 121-742, Korea }

\author{Hideki Maeda \footnote{E-mail address: h-maeda@hgu.jp}}
\affiliation{ Department of Electronics and Information Engineering, Hokkai-Gakuen University, Sapporo 062-8605, Japan.}
\date{\today}

\begin{abstract}
We study causal properties of the recently found rotating black-hole solution in the low-energy
sector of Ho\v{r}ava gravity as a viable Lorentz-violating (LV) gravity in four dimensions
with the LV Maxwell field and a cosmological constant $\Lambda (>-3/a^2)$ for an arbitrary rotation parameter $a$. The region of non-trivial causality violation containing closed timelike curves is
exactly the same as in the Kerr-Newman or the Kerr-Newman-(Anti-)de Sitter solution. Nevertheless,
chronology is protected in the new rotating black hole because the causality violating region becomes
physically inaccessible by exterior observers due to the new {\it three-}curvature singularity at
its boundary that is topologically two-torus including the usual ring singularity at
$(r,\theta)=(0,\pi/2)$. As a consequence, the physically accessible region outside the torus
singularity is causal everywhere.
\end{abstract}


\maketitle

\newpage

\section{Introduction}

It is known that the Kerr family of solutions in general relativity (GR) \cite{Kerr:1963}
suffers from non-trivial causality violation due to closed timelike curves (CTCs) in the
region $r<0$ near the ring singularity in the Boyer-Lindquist coordinates~\cite{Cart:1968}. Moreover, there is no physical obstacles in GR for deforming CTCs to pass any point
inside the inner horizon ($-\infty<r<r_-$). Hawking has proposed the chronology protection
conjecture asserting that the law of physics prevents the appearance of CTCs~\cite{Hawk:1991},
however, it remains unproven at present.

For this problem, it has been argued that chronology inside a Kerr black hole can be
recovered by the dynamical instability. In fact, it has been shown both numerically and
analytically~\cite{Ori:1992zz} that the inner horizon $r=r_-$ possesses a mass inflation
instability~\cite{Poisson:1990eh}, resulting in the appearance of a curvature singularity at
$r=r_-$ that prevents exterior observers from reaching the region where CTCs exist, though its
mathematically rigorous proof beyond the linear level is still absent~\cite{Gurriaran:2024epv}.
As another point of view, it is unclear whether there exists a priori reason to prohibit the
causality violation in the small scale structure of space-time at the Planck scale, where
violent space-time quantum fluctuations could happen, as argued first by Wheeler~\cite{Whee:1957}.
It is therefore interesting to see how these results in GR will be changed in a modified theory
of gravity realized in the low-energy limit of quantum gravity.

A complete formulation of quantum gravity has not yet been achieved.
In this context, Ho\v{r}ava gravity has been proposed as a renormalizable (quantum) gravity without the ghost problem through the $z$-order higher-spatial-derivatives with anisotropic scaling dimension $z$ ($z=3$ in four dimensions), while keeping the second-order time derivatives in the
action, which break the Lorentz symmetry apparently. At present, regrettably, exact rotating
black-hole solutions are not available in the renormalizable {\it full} Ho\v{r}ava gravity in
four dimensions although a massless rotating solution~\cite{Park:2023} and three-dimensional rotating black-hole solutions~\cite{Park:2012, Soti:2014} have been obtained. In these circumstances, an exact rotating black-hole solution has recently been obtained in Ref.~\cite{Deve:2024} in the low-energy sector of non-projectable Ho\v{r}ava gravity~\cite{Hora:2009} as a viable Lorentz-violating (LV) gravity in four dimensions with the LV Maxwell field and with or without a cosmological constant $\Lambda$. Some basic properties of the geometry, like black hole thermodynamics, horizons, and geodesic structure, were studied in Ref.~\cite{Deve:2024}. However, the uniqueness of our solution and its stability are still open problems.

In this letter, we study causal properties of the rotating black-hole solution obtained in
Ref.~\cite{Deve:2024} in the low-energy sector of non-projectable Ho\v{r}ava gravity. The region
of {\it non-trivial} (irremovable by taking a covering space~\cite{Cart:1968}) causality
violation containing CTCs in this solution is exactly the same as in the Kerr-Newman or
the Kerr-Newman-(Anti-)de Sitter solution. Nevertheless, we will show that chronology is
protected in the new LV rotating black hole since the causality violating region becomes
physically inaccessible due to a new {\it three}-curvature singularity at its outer boundary,
which is topologically torus and prevents one reaching to or coming out of it. As a consequence,
the region outside the torus singularity is causal everywhere.

\section{Rotating black hole in the Low-energy sector of non-projectable Ho\v{r}ava gravity}
\label{sec:main}

In Ref.~\cite{Deve:2024}, a stationary and axisymmetric solution was obtained in the low-energy sector of non-projectable Ho\v{r}ava gravity in four dimensions \cite{Hora:2009} with the
LV Maxwell action and a cosmological constant $\Lambda$. (Hereafter, we shall use the unit $c=1$
unless otherwise stated.) The metric and gauge potential of the solution are given in the
Boyer-Lindquist coordinates as \footnote{In this paper, we use the metric signature $(-,+,+,+)$
and we adopt conventions for curvature tensors as
$[\nabla _\rho ,\nabla_\sigma]V^\mu ={{R}^\mu }_{\nu\rho\sigma}V^\nu$ and
${R}_{\mu \nu }={{R}^\rho }_{\mu \rho \nu }$, where Greek indices such as $\mu$ or $\nu$ run over
all spacetime indices, while Latin indices such as $i$ and $j$ run from $1$ to $3$. }
\beq
\D s^2=-N^2 
\D t^2+\frac{\rho^2}{\Delta_r (r)}\D r^2+ \frac{\rho^2}{\Delta_{\theta} (\theta)} \D \theta^2+\frac{\Sigma^2 \sin^2\theta}{\rho^2 \Xi^2 }\left(\D \phi+ N^{\phi} \D t \right)^2,
\label{KAdS_ansatz}
\eeq
and
\begin{align}
A_\mu \D x^\mu=&-\sqrt{\f{\xi}{\eta}} \f{q_e r \De_\theta + q_m a \cos\theta~ (1-{\La r^2/3})}{\rho^2 \Xi} \D t \nonumber\\
&+ \f{1}{\sqrt{\eta \ka}} \f{q_e r {\it a}~ \sin^2\theta+q_m (r^2+a^2)\cos\theta}{\rho^2 \Xi} \D \phi,
\label{A_sol}
\end{align}
respectively, where
\begin{align}
\label{KAdS_sol}
\begin{aligned}
{\rho}^2 =& r^2 + a^2 \cos^2 \theta, \\
{\Delta}_r =& ( r^2 + a^2 ) \left(1 -{\f{\La r^2}{3}}\right)-2mr+q_e^2+q_m^2, \\
{\Delta}_\theta =& 1+\f{\La a^2 \cos^2 \theta}{ 3},\qquad {\Xi}=1+{\f{\La a^2 }{ 3}}, \\
{\Sigma}^2 =& ( r^2 + a^2 ) \rho^2 \Xi+ ( 2mr-q_e^2-q_m^2 )a^2 \sin^2\theta \\
=& {( r^2 + a^2 )^2 {{\Delta}_\theta} -{\Delta}_r a^2 \sin^2\theta}  ,  \\
N^2=&\frac{\rho^2 \Delta_r \De_\theta}{\Sigma^2},\qquad N^{\phi}=-\f{ a ( 2mr-q_e^2-q_m^2 ) {\Delta}_\theta \sqrt{\kappa \xi} }{\Si^2}.
\end{aligned}
\end{align}
Non-vanishing components of the inverse metric $g^{\mu\nu}$ and the determinant of the metric $g\equiv \det(g_{\mu\nu})$ are given by
\begin{align}
\label{inverse-g}
\begin{aligned}
&g^{tt}=-N^{-2},\qquad g^{t\phi}=N^\phi N^{-2},\qquad g^{rr}=\Delta_r \rho^{-2},\qquad g^{\theta\theta}=\Delta_\theta\rho^{-2},\\
&g^{\phi\phi}=\frac{N^2\rho^2\Xi^2-\Sigma^2(N^\phi)^2\sin^2\theta}{\Sigma^2 N^2\sin^2\theta},\qquad g=-\frac{\rho^4\sin^2\theta}{\Xi^2}.
\end{aligned}
\end{align}
Domains of the angular coordinates are given by $\theta\in(0,\pi)$ and $\phi\in[0,2\pi)$ as $\theta=0$ and $\pi$ are coordinate singularities. Since the zeros of ${\Delta}_r(r)$ are also coordinate singularities, if there are, the coordinate system (\ref{KAdS_ansatz}) covers multiple domains of $r$ separately.

The solution is parameterized by the mass parameter $m$, rotation parameter $a$, and electric and
magnetic charges $q_e$ and $q_m$. Other constants $\ka$, $\la$, $\xi$, $\eta$, and $\zeta$ are
coupling constants in the following action\footnote{The solution is valid for an arbitrary $\la$ due to $K=0$, {\it i.e.}, ``maximal slicing".}
\begin{align}
S= \int_{{\bf R} \times \Si_t} \D t \D^3 x
\sqrt{g}N\left[\frac{1}{\kappa}\left(K_{ij}K^{ij}-\lambda
K^2\right)+\xi {(R^{}-2 \La)} 
-\f{2 \eta}{N^2} \left(E_i +F_{ij}N^j \right)^2+\zeta F_{ij} F^{ij}\right]{.}
\label{horava}
\end{align}
Here,
$
K_{ij}=({2N})^{-1}\left(\dot{g}_{ij}-\nabla_i
N_j-\nabla_jN_i\right)
$
is the extrinsic curvature of the time-slicing hypersurface $\Si_t$ given by $t=$constant,
$R$ is the {\it three}-scalar curvature on $\Si_t$, and $E_i=\dot{A}_i -\nabla_i A_0$ and
$F_{ij}=\nabla_i A_j-\nabla_j A_i$ are electromagnetic field-strength three-tensors, where
a dot denotes the time derivative and $\nabla_i$ is the covariant derivatives on $\Si_t$ with
the induced metric $g_{ij}$. Under the ``noble" condition $\zeta \eta^{-1}=\ka \xi$, the speed
of gravitational wave $c_g$ is identical to the speed of light $c_l=c_g=\sqrt{\ka \xi}$ but
not necessarily to the speed of light in vacuum $c{(=1)}$.

Note that the solution given by Eqs.~(\ref{KAdS_ansatz})--(\ref{KAdS_sol}) admits
only the parameter region $\kappa \xi>0$, which we assume throughout this letter, in order that
the solution is real valued, though it is not constrained by the theory (\ref{horava}) itself.
Moreover, in the presence of a negative cosmological constant $\Lambda<0$,  we need to
restrict $\La>-3/a^2$ so that $\Xi>0$ and $\De_{\theta}>0$ always hold in order to avoid a
bizarre spacetime. For example, in the case of $\Xi<0$, the metric admits a non-Lorenzian signature $(-,+,-,-)$ in the region with $\De_{\theta}<0$ and $\De_r>0$, due to $\Sigma^2<0$ and $N^2>0$ by Eq.~(\ref{KAdS_sol}), whereas the metric in the region with $\De_{\theta}>0$ and $\De_r>0$ retains a Lorenzian signature $(-,+,+,+)$.

Depending on the parameters, the solution given by Eqs.~(\ref{KAdS_ansatz})--(\ref{KAdS_sol})
describes a charged rotating black hole. In the GR limit $\ka{\to} \xi^{-1}$, the solution
reduces to the Kerr-Newman-(Anti-)de~Sitter solution and $c_g=c_l=1$ is
satisfied~\cite{Kerr:1963,Cart:1968}. However, for $\ka \neq \xi^{-1}$, we have $c_g=c_l \neq 1$
and there are sharp non-trivial LV effects as will be discussed below. Hereafter, we consider the
simplest case without a cosmological constant nor electro-magnetic charges, {\it i.e.}, $\La=0$
and $q_e=q_m=0$. However, our main conclusion is unchanged even in the most general case with
$q_e\ne 0$, $q_m\ne 0$, and ${\La>-3/a^2}$.

With arbitrary values of the parameters $\ka$, $\la$, $\xi$, $\eta$, and $\zeta$, the {\it apparent} gravitational symmetry of the action is the ``foliation-preserving" diffeomorphism
(\DiffF)~\cite{Hora:2009,Park:2009}, and the physical singularities are captured by
\DiffF~-invariant curvatures. Up to finite factors, such \DiffF~-invariant curvatures for our solution (\ref{KAdS_ansatz})--(\ref{KAdS_sol}) are given by $K=0$, $K_{ij} R^{ij}=0$, and
\beq
R \sim \f{a^2 m^2}{\rho^6 \Si^4},\qquad R_{ij} R^{ij} \sim \f{m^2}{\rho^{12} \Si^8},\qquad K_{ij} K^{ij} \sim \f{\ka \xi a^2 m^2}{\rho^6 \Si^4}.
\label{3D_curvature}
\eeq
Equation~(\ref{3D_curvature}) shows that the solution admits a curvature singularity \footnote{This new singularity is defined as a true singularity in our {\it non-projectable}-foliation solution. However, it remains an interesting open question whether the new singularity structure is altered in a projectable-foliation solution. The projectable solution is considered to be physically distinct because it cannot be transformed from our non-projectable solution via ${\it Diff}_{\cal F}$.} determined by
$\Si^2= ( r^2 + a^2 ) \rho^2 +2mr a^2 \sin^2\theta=0$, as well as the usual ring singularity at
$(r,\theta)=(0,\pi/2)$ determined by $\rho^2=0$ in the Kerr solution.

\begin{figure}
\includegraphics[width=8cm,keepaspectratio]{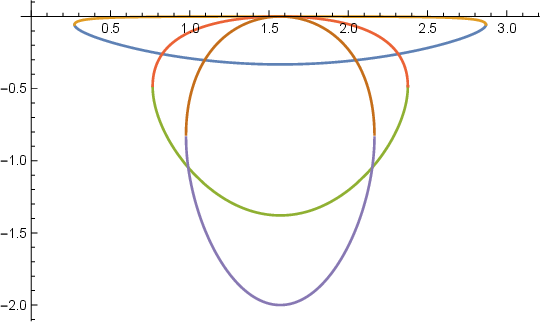}
\caption{\label{Fig:singularity-torus} $r$ vs. $\theta{\in[0,\pi]}$ of singularity surfaces
determined by $\Si^2 (r, \theta)=0$ for $m=2$ with $a=2$, $1$, and $0.1$, which correspond to
the closed curves from the bottom to the top, respectively. As $a \ra 0$, the singularity
surfaces reduce to a point singularity of the \Sch black hole at $r=0$.}\label{fig:singular}
\end{figure}

The singularity of $\Si^2=0$ appears in the region where $\Delta_r>0$ and $mr<0$ hold, and it
is given by
\beq
\sin^2\theta=\f{(r^2+a^2)^2}{a^2 {[}-2 mr +(r^2+a^2)\bc{]}}
\eeq
with $\phi\in[0,2\pi)$, which includes the usual ring singularity at $(r,\theta)=(0,\pi/2)$.
We consider a spacetime region which admits an asymptotically flat end $r\to +\infty$.
Then, as we will see that the domain of $r$ is given by $-\infty<r<\infty$ and the solution is invariant under $(m,r)\rightarrow (-m,-r)$, we can take $m>0$ without loss of
generality. For $m>0$, as shown in Fig.~\ref{Fig:singularity-torus}, this singularity
is located in the region $r\le 0$ and topologically {\it two}-torus. The torus surface does not depend on the LV parameter $\ka \xi$ and keeps the same form in both the LV and GR cases.

The new singularity belonging to the torus is independent of the ring singularity and the sharp appearance of the new singularity of $\Si^2=0$ with the LV parameter
$\ka \neq \xi^{-1}$, in contrast to the usual ring singularity of $\rho^2=0$ in the GR case $\ka= \xi^{-1}$, can be clearly seen in the four-dimensional curvature invariants,
\begin{align}
\label{4D_curvature}
\begin{aligned}
&R^{(4)} \sim (\ka \xi-1)~ \f{a^2 m^2}{\rho^6 \Si^4},\qquad R^{(4)}_{\m \n} R^{(4) \m \n} \sim (\ka \xi-1)^2~ \f{a^4 m^4}{\rho^{12} \Si^8}, \\
&R^{(4)}_{\m \n \si \rho} R^{(4) \m \n \si \rho} \sim (\ka \xi-1)~ \f{m^2}{\rho^{12} \Si^8}+\left(\cdots\right)\f{m^2}{\rho^{12}},
\end{aligned}
\end{align}
where $\left(\cdots\right)$ is the same factor as in GR. Eq.~(\ref{4D_curvature}) shows that
the new singularity of $\Si^2=0$ disappears discontinuously and only the usual ring singularity of $\rho^2=0$
is left in the GR case $\ka= \xi^{-1}$, {\it i.e.}, the Kerr solution, for which physical quantities are the four-dimensional {\it Diff}-invariant ones shown in Eq.~(\ref{4D_curvature}), not the three-dimensional \DiffF~-invariant ones shown in Eq.~(\ref{3D_curvature}). We also note that the torus singularity is more severe than the usual ring singularity generally because the torus singularity produces extra divergence on top of the ring singularity, if exists, as can be seen in Eq. (\ref{4D_curvature}).

In the Kerr-Newman-(Anti-)de~Sitter solution in GR described by the metric (\ref{KAdS_ansatz}), $r=r_{\rm h}$ defined by $\Delta_r(r_{\rm h})=0$ is a Killing horizon. The regularity of $r=r_{\rm h}$ and the extension of spacetime beyond there are transparently shown in the horizon-penetrating coordinates such as the Doran coordinates~\cite{Doran:1999} or Kerr's original coordinates~\cite{Kerr:1963}. (See Ref.~\cite{Visser:2007fj} for a review.)
We can define the future direction consistently on both sides of the Killing horizon in such
horizon-penetrating coordinates, whereas it is not possible in the Boyer-Lindquist coordinates
(\ref{KAdS_ansatz}). It is a non-trivial problem in our theory (\ref{horava}) if there exist
horizon-penetrating coordinates obtained by \DiffF~-invariant coordinate transformations from
the Boyer-Lindquist coordinates (\ref{KAdS_ansatz}). For this reason, we will study causal
properties of our solution independent from the definition of the future direction.

\section{Chronology protection by the torus singularity}

\begin{figure}
\includegraphics[width=15cm,keepaspectratio]{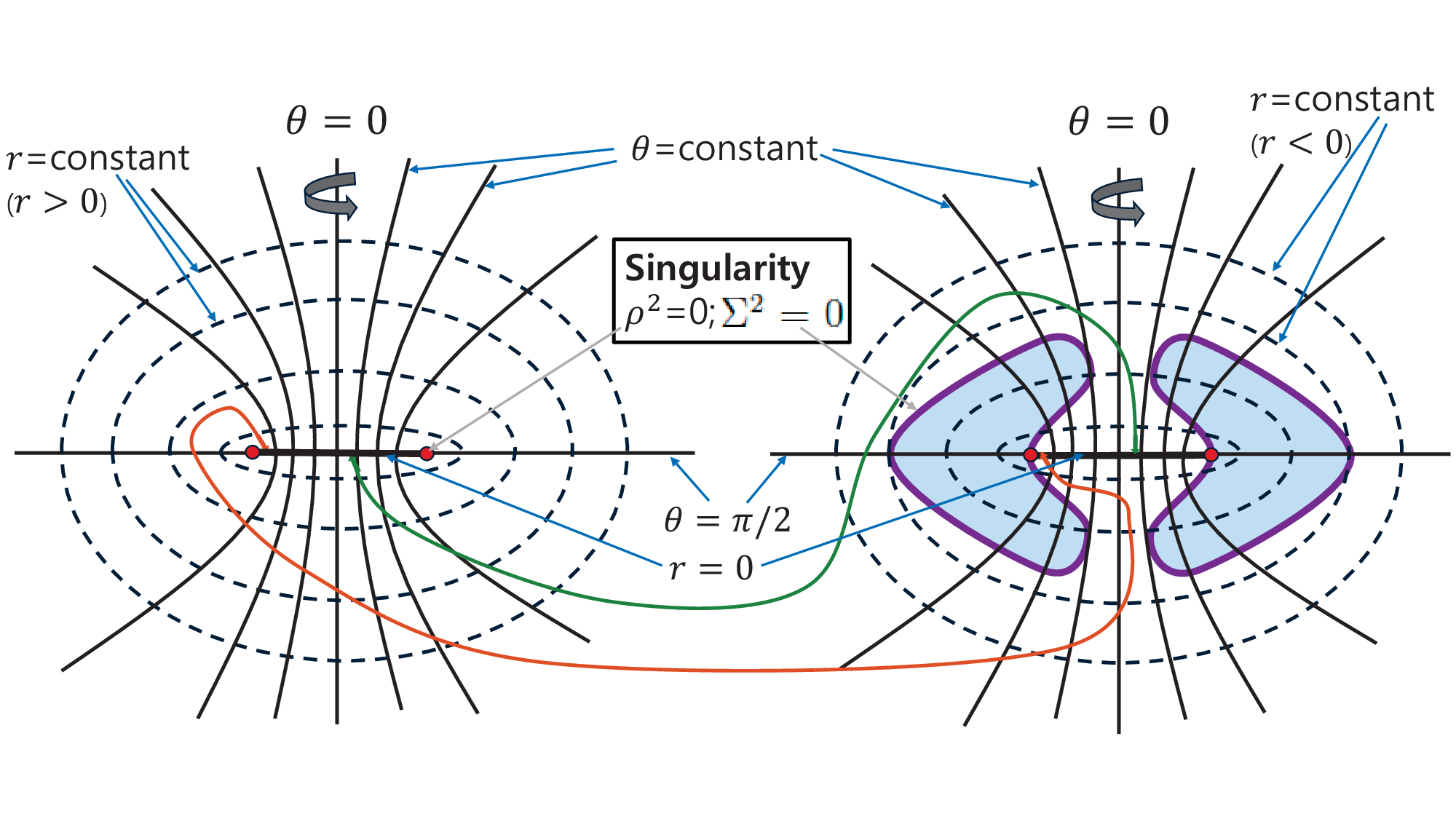}
\caption{\label{Fig:CTC_Kerr}The maximal extension of the rotating black-hole solution
for $m^2>a^2$ by identifying the top of the disk $x^2+y^2<a^2$ with $z=0$ in the region $r>0$
(the left chart) with the bottom of the corresponding disk in the region $r<0$ (the right chart)
and vice versa. The torus singularity of $\Si^2=0$ exists in the region $r\leq 0$ that includes
the usual ring singularity of $\rho^2=0$ located at $(r,\theta)=(0,\pi/2)$.}
\label{fig:CTC}
\end{figure}

In order to study causal structure of our rotating black hole spacetime described by
the metric (\ref{KAdS_ansatz}) with $m>0$, $\La=0$, and $q_e=q_m=0$, we first consider
the case $m^2>a^2$, in which there are two Killing horizons at $r=r_\pm:=m\pm\sqrt{m^2-a^2}$
determined by
\beq
\De_r =r^2+a^2-2mr=0,
\eeq
which are exactly the same as in the Kerr solution. As in the Kerr black hole~\cite{Hawk:1973},
$r=0$ of our rotating black hole is not the end of the spacetime but it can be extended
regularly into the region $r<0$ through the interior of the disk defined by $x^2+y^2<a^2$ with
$z=0$ in the Cartesian coordinates $x=(r^2+a^2)^{1/2} \sin\theta\cos \phi$,
$y=(r^2+a^2)^{1/2}\sin\theta\sin\phi$, and $z=r\cos\theta$~{\cite{Hawk:1973,ONei:2014}, which are
\DiffF~-invariant transformations. As shown in Fig.~\ref{Fig:CTC_Kerr}, if a regular coordinate
system covering the Killing horizons is admitted, the maximally extended spacetime of our black
hole is given in the domains $r\in (-\infty,+\infty)$, $\theta\in(0,\pi)$, and $\phi\in[0,2\pi)$
satisfying $\Sigma^2>0$ due to the torus singularity of $\Si^2=0$}.

Then, in this spacetime, closed timelike curves (CTCs) exist in the region where $g_{\phi \phi}<0$
holds so that a vector $\pa_{\phi}$ becomes timelike, violating causality. The causality
violating region ${\cal V}=\{ g_{\phi \phi}<0 \}$ of the solution, determined by $\Sigma^2<0$,
is located in the region $r<0$ and exactly the same as in the Kerr solution due to the same
form of $g_{\phi \phi}$~\cite{Cart:1968}. However, the important difference of our LV rotating
black-hole solution from Kerr is that the torus singularity of $\Si^2=0$ prevents us from
reaching the causality violating region ${\cal V}$ from the usual regions where $\Si^2>0$ holds,
{\it e.g.}, from region III$'$, by dividing the regions as I $=\{r>r_+\}$, ~II $=\{r_-<r<r_+\}$,~III $=\{0<r<r_-\}$, and~III$'$ $=\{ r{\le}0\}-{\cal V}$.
Moreover, there is no way of deforming CTCs in the causality violating region ${\cal V}$ to
pass any point of region III$'$ {because region III$'$ is} protected by the torus singularity of
$\Sigma^2=0$ (cf. Ref.~\cite{Cart:1968}). On the other hand, all of regions I, II, and
III$\cup$III$'$ outside the causality violating region ${\cal V}$ are casually well behaved
as shown in Ref.~\cite{Cart:1968} and Proposition 2.4.6 in Ref.~\cite{ONei:2014} for
the Kerr spacetime. (We recap the proof in Ref.~\cite{ONei:2014} in Appendix~\ref{app:ONeil}.)

For $m^2=a^2$, the two horizons $r_+$ and $r_-$ coincide and the region II disappears, where
$\Delta_r<0$ holds. However, other regions I and III$\cup$III$'$ remain and the above result
on their causality non-violation is still valid even though the causality violating region
${\cal V}$ is more elongated to the equator in {the region} $r<0$ such as the bottom curve
in Fig.~\ref{Fig:singularity-torus}.

For $m^2<a^2$, we have $\Delta_r>0$ everywhere, and there are only regions I$'$ $=\{r>0\}$ and
III$'$ $=\{ r{\le}0\}-{\cal V}$ with no horizons and the torus singularity is globally naked.
The maximally extended region consisting of regions I$'$ and III$'$, outside the causality violating region ${\cal V}$, are casually well behaved, for the same reason as in the case of $m^2 > a^2$.

A time-orientable spacetime is said to be {\it causal (chronological)} if there is no
closed causal (timelike) curve~\cite{Hawk:1973}. Thus, there is a {\it chronology protection} in
our rotating solution due to the torus singularity of $\Si^2=0$ at the outer boundary of the
causality violating region ${\cal V}$ when there is the Lorentz violation with
$\ka \neq \xi^{-1}$. We finally note that our conclusion remains valid even when we consider
the generalized rotating solution with the electromagnetic charges $q_e$ and $q_m$ and a
cosmological constant $\La$ satisfying $\La>-3/a^2$ so that the metric retains Lorentzian
signature for all $\theta\in(0,\pi)$. In Appendix~\ref{app:generalization}, we discuss the
behavior of the singularity surface of $\Sigma^2=0$, the causality-violating region in the limit $\La\to -3/a^2$, and the bizarre behaviors beyond that limit.

If there exists a {\it time function} whose gradient is timelike\footnote{If a time function is valid in the entire spacetime, it is referred to as a {\it global} time function.},
one can greatly strengthen our discussions on causality~\cite{Hawk:1973,Wald:1984}.
A time-orientable spacetime is said to be {\it stably causal} if no CTC appears even under
any small deformation against the metric~\cite{Hawk:1973}. By Proposition 6.4.9 in Ref.~\cite{Hawk:1973}, a spacetime region is stably causal if and only if there exists a time function. The following proposition shows a chronology protection in our rotating solution in
the most general case with any values of $m$, $q_e$, $q_m$, and $\La(>-3/a^2)$. \\

\noindent
{\bf Proposition~1}: {\it A maximally extended spacetime of the solution described by
Eqs.~(\ref{KAdS_ansatz})--(\ref{KAdS_sol}) with $\La ~>-3/a^2$ is causal.}\\

\noindent
{\it Proof}.
Due to the torus singularity of $\Si^2=0$, the maximally extended spacetime of the solution is given in the domains $r\in (-\infty,+\infty)$, $\theta\in(0,\pi)$, and $\phi\in[0,2\pi)$ satisfying $\Sigma(r,\theta)^2>0$. In addition, $\De_\theta>0$ is satisfied for $\La >-3/a^{2}$.
The regions where $\Delta_r(r)>0$ holds are stably causal because $T=\pm t$ is a time function, shown by $(\nabla_{\mu} T)( \nabla^{\mu} T)=-\Sigma^2/(\rho^2\Delta_r \De_{\theta})<0$.
The regions where $\Delta_r<0$ holds are also stably causal because $T=\pm r$ is a time
function, shown by $(\nabla_{\mu} T)( \nabla^{\mu} T)=\Delta_r/\rho^2<0$. Here the signs in the
definitions of $T$ are chosen such that $T$ increases in the future direction. Since the regions
with $\Delta_r\ne 0$ are stably causal, the only possibility to have CTCs in the maximally
extended spacetime is that the turning points along the CTCs are located at the horizons defined
by $\Delta_r(r_{\rm h})= 0$. If $r=r_{\rm h}$ is a turning point of a CTC, the CTC must be
tangent to a null hypersurface $r=r_{\rm h}$. However, it is not possible because the tangent
vector of the CTC is timelike, whereas independent tangent vectors of a null hypersurface
consist of a null vector and two spacelike vectors. $\Box$\\


The difference of the time function $T$ reflects the fact that $t$ and $r$ are timelike
coordinates in the regions $\Delta_r>0$ and $\Delta_r<0$, respectively, and hence our proof is
similar to that in Ref.~\cite{Cart:1968} or Proposition 2.4.6 in Ref.~\cite{ONei:2014}. Actually,
our proof improves Proposition 2.4.6 in Ref. \cite{ONei:2014} that shows causality only in the
regions away from the horizons, as recapped in Appendix~\ref{app:ONeil}.

\section{Concluding remarks}

In this letter, we have studied causal properties of the rotating black-hole solution given by
Eqs.~(\ref{KAdS_ansatz})--(\ref{KAdS_sol})~\cite{Deve:2024} in the low-energy sector
of non-projectable Ho\v{r}ava gravity~\cite{Hora:2009} as a viable Lorentz-violating (LV)
gravity in four dimensions with the LV Maxwell field and a cosmological constant
$\La(>-3/a^2)$. In spite that the region of causality violation containing CTCs in this
solution is exactly the same as in the Kerr-Newman or the Kerr-Newman-(Anti-)de~Sitter
solution, we have shown in Proposition~$1$ that the maximally extended spacetime of this new solution is {\it causal} everywhere including horizons because the causality violating region becomes physically inaccessible due to the torus singularity at the boundary of causality violating region ${\cal V}=\{ g_{\phi \phi}<0 \}$ with the Lorentz violation $\ka \neq \xi^{-1}$.
The present result supports Hawking's conjecture on the existence of ``the law of physics" that
protects chronology~\cite{Hawk:1991}.

In spite that the horizons determined by $\Delta_r(r)=0$ are {\it coordinate} singularities in
the Boyer-Lindquist coordinates (\ref{KAdS_ansatz}), we have shown in Proposition $1$ that
there is no CTC everywhere including horizons by constructing time functions defined in
spacetime regions covered by the coordinates (\ref{KAdS_ansatz}). Then, one might think that
we can even prove that the whole spacetime, including horizons, is {\it stably} causal by
constructing a {\it global} time function in the Doran-like horizon-penetrating coordinates
covering the horizons~\cite{Doran:1999}. In GR, the Kerr vacuum solution can be described in the Doran coordinates~\cite{Doran:1999} which cover the region $r\ge 0$ including the horizons $r=r_\pm$ for $m>0$. Moreover, as we have $g^{\tau\tau}=-1$ with the Doran time coordinate
$\tau:= t +\int^r_{0} \sqrt{2mr (r^2+a^2)}/\De_r \D r$, $T(\tau,r)= \tau$ is a global time
function in this region that satisfies $(\nabla^{\mu} T)(\nabla_{\mu} T)=-1$ and, as a
consequence, we can prove that the region in the Kerr spacetime where $\Sigma^2>0$ holds
including the horizons $r=r_\pm$ is stably causal. However, because
$t \ra \tau-\int^r_{0} \sqrt{2mr (r^2+a^2)}/\De_r \D r$ is {\it not} a symmetry
transformation in the LV action (\ref{horava}), we can {\it not} obtain the Doran-like solution
from the Kerr-like solution given by Eqs.~(\ref{KAdS_ansatz})--(\ref{KAdS_sol})
by simply replacing $t$ into ${\tau}$ and we need to find a {\it Doran-like}
rotating black-hole solution separately~\cite{Doran-like}.

Lastly, to discover rotating black-hole solutions in the renormalizable {\it full} Ho\v{r}ava
gravity is surely an important outstanding problem. We may expect that higher-derivative
Lorentz-violating terms can make curvature singularities milder due to
{\it non-perturbative effects} than those without higher-derivative
terms~\cite{Lu:2009, Keha:2009, Park:0905, Kiri:2009} or produce additional curvature
singularities~\cite{Cai:2010, Argu:2015, Park:2012, Soti:2014}. However, it is quite
questionable whether the new torus singularity at the low-energy is {\it completely}
removed by the non-perturbative higher-derivative effects {so that the chronology protection
disappears in the rotating black-hole solution for the {\it full} Ho\v{r}ava gravity. This problem is left for future investigation.

\section*{Acknowledgments}

This work was supported by Basic Science Research Program through the National
Research Foundation of Korea (NRF) funded by the Ministry of Education,
Science and Technology (RS-2020-NR049598). The authors thank the organizers of the conference ``String theory, Gravity and Cosmology 2024 (SGC2024)'', held at the Institute for Basic Science (IBS) in Daejeon, Korea, on December 4--7, 2024, where this work was initiated.

\appendix

\section{ Recap of Proposition 2.4.6 in Ref.~\cite{ONei:2014}}
\label{app:ONeil}

In this appendix, we recap the proof of Proposition 2.4.6 in Ref.~\cite{ONei:2014} for the Kerr spacetime with $m>0$, $q_e=q_m=0$, and $\La=0$.\\

\noindent
{\bf Proposition~2.4.6 (\protect{O'Neill~\cite{ONei:2014}})}: {\it For $m^2\ge a^2$, regions {\rm I}, {\rm II}, and {\rm III}$\cup${\rm III}$'$ are causal.}\\

\noindent
{\it Proof}. A spacetime ${\cal M}$ is causal (chronological) if there are no closed non-spacelike (timelike) curves in ${\cal M}$. In order to prove the proposition, we first show that the hypersurface ${\cal N}$ of $t=constant$ is {\it spacelike} in regions I and III$\cup$III$'$.
To this end, we note that any tangent vector $\boldsymbol{v}$ at each point $p \in {\cal N}$ can be written as
\beq
{\boldsymbol{v}}=v^r \pa_r+v^\theta \pa_\theta+v^\phi \pa_\phi
\eeq
with the mutually orthogonal basis vector fields $\pa_r$, $\pa_\theta$, and $\pa_\phi$ that
span the target space $T_p ({\cal N})$. Then, we have
\beq
{\boldsymbol{v}}\cdot {\boldsymbol{v}}=(v^r)^2 g_{rr} +(v^\theta )^2 g_{\theta \theta}+(v^\phi)^2 g_{\phi \phi} >0
\eeq
since $g_{rr}, g_{\theta \theta}$, and $g_{\phi \phi}$ are positive. Hence, the hypersurface
${\cal N}$ of $t=constant$ is {\it spacelike}. This is equivalent to the fact that its normal
vector $\nabla^{\mu} t$ is timelike, shown by $(\nabla^{\mu} t)(\nabla_{\mu} t)=g^{tt}=-\Si^2/(\rho^2 \De_r)<0$.

We next show by contradiction that, along any non-spacelike $C^1$ curve $x^{\mu} (\la)$
parameterized by $\la$, the coordinate $t(\la)$ is strictly monotonic and therefore we can set
$\lambda$ such that $\D t(\la)/\D\la > 0$ without loss of generality using the degree of
freedom $\lambda\to -\lambda$. Suppose that there exists $\la=\la_1$ satisfying
$
\D t/{\D \la}{|_{\lambda=\lambda_1}}=0
$.
Then, $v^{\mu}=\D x^{\mu}/\D\la{|_{\lambda=\lambda_1}}$ is tangent to a hypersurface
$t=t_1 (\la_1)$, which gives a contradiction because $\D x^{\mu} /\D \la$ is {\it not}
spacelike by the assumption that the curve is non-spacelike, whereas we have shown in the
above that the $t=constant$ hypersurface ${\cal N}$ is spacelike. This proves the proposition for
regions I and III$\cup$III$'$ since a {\it closed} non-spacelike curve needs at least one
spacetime point {\it p} where $\D t/\D \la|_{\it p} = 0$ holds. (See Fig.~\ref{fig:curve}.)

\begin{figure}
\includegraphics[width=13cm,keepaspectratio]{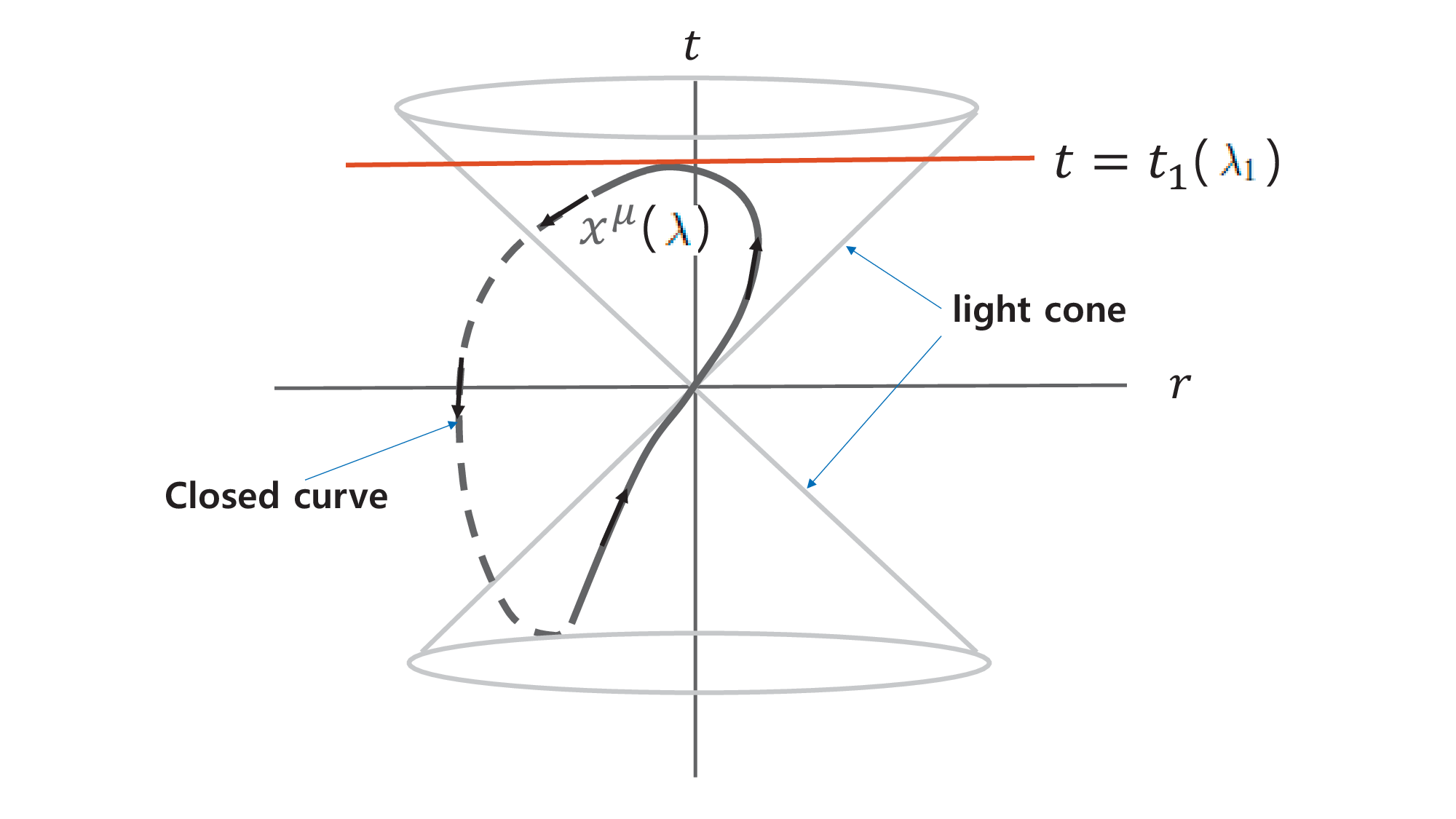}
\caption{A closed curved in the spacetime, which cannot be non-spacelike everywhere.}
\label{fig:curve}
\end{figure}

In region II, $\pa_r$ is timelike and the hypersurface ${\cal N}$ of $r=constant$,
whose tangent space is spanned by $\pa_t$, $\pa_\theta$, and $\pa_\phi$, is spacelike,
which is equivalent to the fact that its normal vector $\nabla^{\mu} r$ is timelike, shown by
$(\nabla^{\mu} r)(\nabla_{\mu} r)=g^{rr}=\Delta_r \rho^{-2}<0$. Then, the same argument for region I or III$\cup$III$'$ works with $t$ and $r$ exchanged such that the time coordinate $r$ is strictly monotonic and hence there is no closed non-spacelike curve in region II as well. $\Box$

\section{Causality violation and singularities in the most general case}
\label{app:generalization}

As Eq.~(\ref{3D_curvature}) shows, the generalized rotating black-hole solution
with the electromagnetic charges $q_e$ and $q_m$ in the presence of a cosmological
constant $\La$ also admits a curvature singularity determined by
\begin{\eq}
\label{singular_AdS}
{\Sigma}^2 = ( r^2 + a^2 ) \rho^2 \Xi+ ( 2mr-q_e^2-q_m^2 )a^2 \sin^2\theta =0,
\end{\eq}
which is solved to give
\begin{\eq}
\sin^2\theta=\f{(r^2+a^2)^2 \left( 1+{\La a^2/3} \right)}{a^2 \left[-2 mr+q_e^2+q_m^2 +(r^2+a^2) \left( 1+{\La a^2/3}\right)\right]}
\label{singular_AdS2}
\end{\eq}
in addition to the usual ring singularity located at $(r,\theta)=(0,\pi/2)$ determined by $\rho^2=0$. Equation~(\ref{singular_AdS}) shows that, regardless of the value of $\Lambda$, the
singularity of ${\Sigma}^2=0$ includes the ring singularity only in the neutral case ($q_e=q_m= 0$). Singularity surfaces (\ref{singular_AdS2}) with different values of the parameters are plotted in Fig.~\ref{fig:singular_general}.

\begin{figure}
\includegraphics[width=7cm,keepaspectratio]{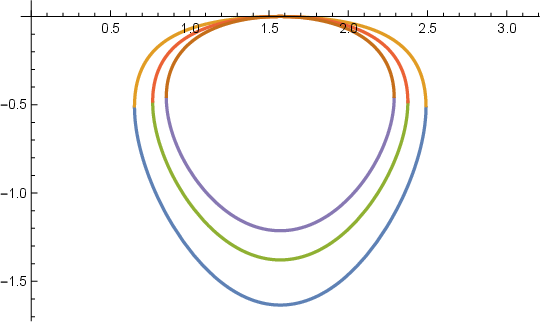}
\qquad
\includegraphics[width=7cm,keepaspectratio]{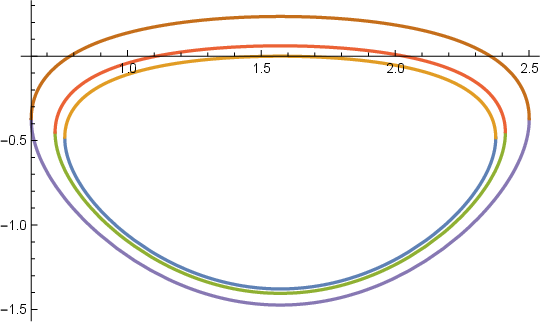}
\caption{$r$ vs. $\theta{\in[0, \pi]}$ of the singularity surface (\ref{singular_AdS2})
with ${\xi=1,}~ a=1$ and $m=2$. In the left panel, we vary $\La=-1,0,1$ (from the outer to
inner curves) with $q_e=q_m=0$. In the right panel, we vary $q_e=0,0.5,1$ (from the inner to
outer curves) with $q_m=0$ and $\La=0$.}
\label{fig:singular_general}
\end{figure}

As seen in the left panel of Fig.~\ref{fig:singular_general}, the role of a cosmological constant
$\La$ in the neutral case ($q_e=q_m= 0$) is just to make either the singularity surface expand
(${\La}<0$) or contract (${\La}>0$). On the other hand, as seen in the right panel of
Fig.~\ref{fig:singular_general}, the role of electromagnetic charges $q_e$ and $q_m$ is to make
the singularity surfaces penetrate into the region} $r>0$ such that there are no overlaps with
the ring singularity at $(r,\theta)=(0,\pi/2)$, and the singularity surface expands as the value
of the charge increases.

\begin{figure}
\includegraphics[width=15cm,keepaspectratio]{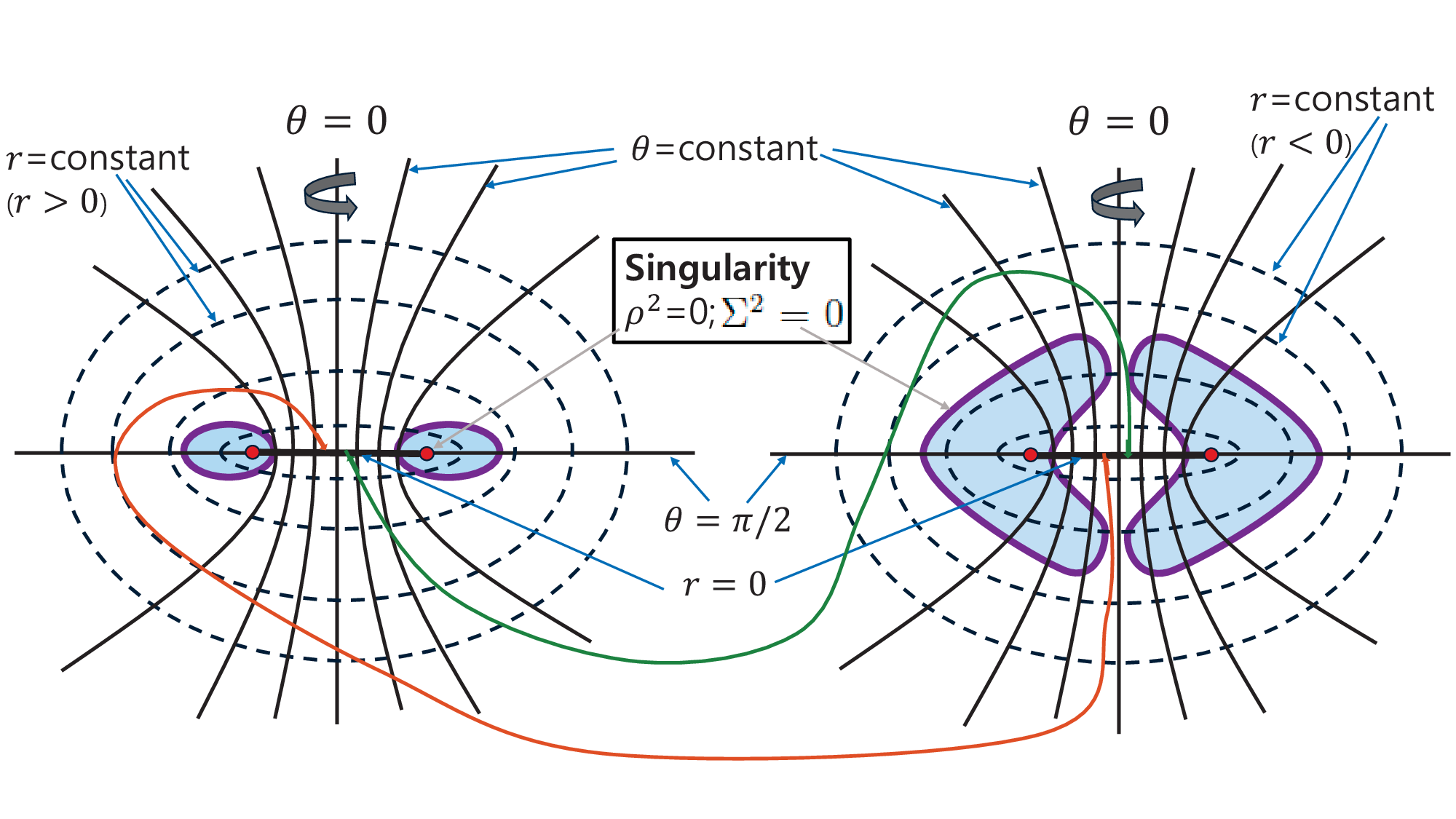}
\caption{The maximal extension of the generalized rotating black-hole solutions with
electromagnetic charges and cosmological constant, with the similar identification of the
disk regions as in Fig. 2. The torus singularity of $\Si^2=0$ spreads in both regions $r>0$ and
$r<0$ and envelopes the ring singularity of $\rho^2=0$ at $(r,\theta)=(0,\pi/2)$.}
\label{fig:CTC_general}
\end{figure}

Note that the torus singularity spreads both in the regions $r\ge 0$ and $r\le 0$ and envelopes
the ring singularity of $\rho^2=0$ at $(r,\theta)=(0,\pi/2)$. (See Fig.~\ref{fig:singular_general}.) By the second expression of $\Sigma^2$ in Eq.~(\ref{KAdS_sol}) with $\La >-3/a^{2}$ or, equivalently, $\Xi>0$ so that $\De_{\theta}>0$ holds for all $\theta\in(0,\pi)$, the penetrated singularity surface in the region $r>0$ does not meet the horizons and is always surrounded by the inner horizon.

\begin{figure}
\includegraphics[width=5.1cm,keepaspectratio]{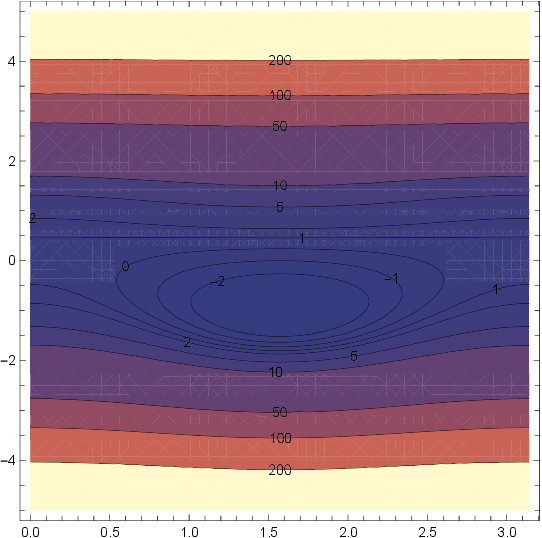}
\quad
\includegraphics[width=5.1cm,keepaspectratio]{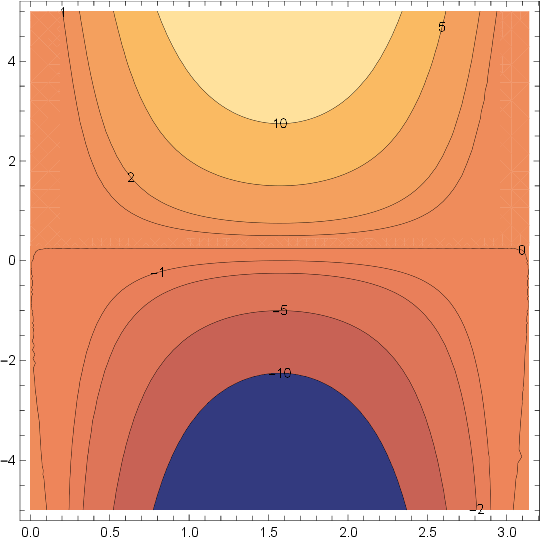}
\quad
\includegraphics[width=5.1cm,keepaspectratio]{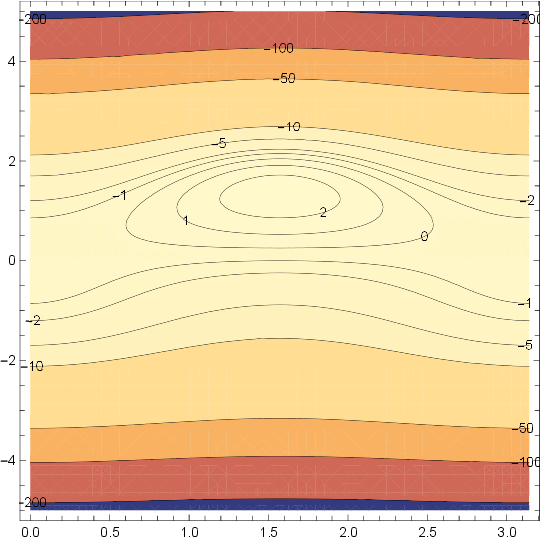}
\caption{$r$ vs. $\theta{\in[0, \pi]}$ contours of $\Si^2 (r, \theta)$ for $-3/a^{2}<\La<0$
(the left and middle panels) and $\La <-3/a^{2}$ (the right panel). We vary $\La=-1, -2.999, -4$
(from the left panel to the right panel) with $a=1$, $m=2$, $q_e=1$, and $q_m=0$.}
\label{fig:contour}
\end{figure}

In this letter, we have assumed $\La >-3/a^{2}$ so that the metric retains the Lorentzian
signature for all $\theta\in(0,\pi)$. Figure~\ref{fig:contour} shows the behavior of the
singularity surface of $\Si^2 (r, \theta)=0$ in the limit $\La \to-3/a^{2}$ and the bizarre
behaviors beyond the limit. When $\La >-3/a^{2}$, the singularity surface of
$\Si^2 (r, \theta)=0$ and the causality-violating region of $\Si^2 (r, \theta)<0$ are located
in the region $r<(q_e^2+q_m^2)/(2m)$ for $m>0$ {as shown in the left panel of
Fig.~\ref{fig:contour}. In the limit of $\La\to -3/a^{2}$}, the singularity surface expands and
approaches a closed curve given by $r=(q_e^2+q_m^2)/(2m)$, $\theta=0$, and $\theta=\pi$, and
finally the causality-violating region becomes the whole lower-half region including
$r\ra -\infty$ surrounded by $r<(q_e^2+q_m^2)/(2m)$, $\theta=0$, and $\theta=\pi$ as shown in
the middle panel of Fig.~\ref{fig:contour}.

The geometry at $\La =-3/a^{2}$ is ill-defined due to infinite determinant $g=-\infty$ in
Eq. (\ref{inverse-g}) (cf. Ref.~\cite{Hawking:1998kw}), but the geometry beyond that critical
point could be still defined. However, as will be discussed below, the geometry beyond that
limit shows the bizarre behaviors causing the non-Lorenzian signature $(-,+,-,-)$.
If $\La <-3/a^{2}$, the causality-violating region extends even to the upper-half region
$r>(q_e^2+q_m^2)/(2m)$ including $r\ra +\infty$ boundary, as well as the lower-half region,
with the contracted singularity surface of $\Si^2 (r, \theta)=0$ and the causal region of
$\Si^2 (r, \theta)>0$, shown in the right panel of Fig. \ref{fig:contour}. On the other hand, for the region $\cos^2 \theta >-3/(\La a^2)$, {\it i.e.}, $\De_{\theta}<0$ near the north and south poles, $\theta=0$ and $\theta=\pi$ with $\De_r>0$, the geometry becomes {\it non-Lorenzian}
with the signature $(-,+,-,-)$ due to $\Sigma^2<0$ and $N^2>0$ by Eq.~(\ref{KAdS_sol}), whereas
the metric for the other region with $\De_{\theta}>0$ retains {\it Lorenzian} signature
$(-,+,+,+)$. In other words, the geometry has {\it both} Lorenzian and non-Lorenzian regions
whose physical relevance seems to be unclear.

In the spacetime described by the metric (\ref{KAdS_ansatz}), we can introduce basis one-forms in the orthonormal frame as
\begin{align}
\begin{aligned}
&e^{(0)}_\mu\D x^\mu=\sqrt{\varepsilon_r\varepsilon_\theta\frac{\rho^2 \Delta_r \De_\theta}{\Sigma^2}}\D t,\qquad e^{(1)}_\mu\D x^\mu=\sqrt{\varepsilon_r\frac{\rho^2}{\Delta_r}}\D r,\\
&e^{(2)}_\mu\D x^\mu=\sqrt{\varepsilon_\theta\frac{\rho^2}{\Delta_{\theta}}}\D \theta,\qquad e^{(3)}_\mu\D x^\mu=\sqrt{\frac{\Sigma^2 \sin^2\theta}{\rho^2 \Xi^2 }}\left(\D \phi+ N^{\phi} \D t \right),
\end{aligned}
\end{align}
which satisfy $g^{\mu\nu}e_{\mu}^{(a)}e_{\nu}^{(b)}=\eta^{(a)(b)}=
\mbox{diag}(-\varepsilon_r\varepsilon_\theta,\varepsilon_r,\varepsilon_\theta,1)$, where $\varepsilon_r:=\mbox{sign}(\Delta_r)$ and $\varepsilon_\theta:=\mbox{sign}(\Delta_\theta)$.
Therefore, the spacetime admits the Lorentzian signature such as $(-,+,+,+)$, $(+,-,+,+)$, and $(+,+,-,+)$ in the regions of $\Delta_r>0$ with $\Delta_\theta>0$, $\Delta_r<0$ with $\Delta_\theta>0$, and $\Delta_r> 0$ with $\Delta_\theta<0$, respectively.
In contrast, the spacetime admits the non-Lorentzian signature $(-,-,-,+)$ in the regions of $\Delta_r<0$ with $\Delta_\theta<0$. The region $\Delta_\theta<0$ appears only for $\La <-3/a^{2}$ in the region where $\cos^2 \theta >-3/(\La a^2)$ holds.

\newcommand{\J}[4]{#1 {\bf #2} #3 (#4)}
\newcommand{\andJ}[3]{{\bf #1} (#2) #3}
\newcommand{\AP}{Ann. Phys. (N.Y.)}
\newcommand{\MPL}{Mod. Phys. Lett.}
\newcommand{\NP}{Nucl. Phys.}
\newcommand{\PL}{Phys. Lett.}
\newcommand{\PR}{Phys. Rev. D}
\newcommand{\PRL}{Phys. Rev. Lett.}
\newcommand{\PTP}{Prog. Theor. Phys.}
\newcommand{\hep}[1]{ hep-th/{#1}}
\newcommand{\hepp}[1]{ hep-ph/{#1}}
\newcommand{\hepg}[1]{ gr-qc/{#1}}
\newcommand{\bi}{ \bibitem}

\end{document}